\documentstyle[aps,prl,twocolumn,epsfig,floats,amssymb]{revtex}

\begin{document}
\draft  

\title{Potential Energy in a Three-Dimensional Vibrated Granular
\\Medium Measured by NMR Imaging}

\author{Xiaoyu Yang and D. Candela}
\address{Physics Department, University of Massachusetts,
Amherst, Massachusetts 01003}

\date{10 January 2000}
\maketitle

\begin{abstract}
	Fast NMR imaging was used to measure the density profile
of a three-dimensional granular medium fluidized by vertical
vibrations of the container.
	For container acceleration much larger than gravity, the
rise in center of mass of the granular medium is found to scale
as $v_0^\alpha/N^\beta$ with $\alpha = 1.0 \pm 0.2$ and
$\beta = 0.5 \pm 0.1$, where $v_0$ is the vibration velocity and
$N$ is the number of grains in the container.
	This value for $\alpha$ is significantly less than found
previously for experiments and simulations in one dimension
($\alpha = 2$) and two dimensions ($\alpha = \text{1.3-1.5}$).
\end{abstract}

\pacs{PACS numbers: 45.70.-n, 81.05.Rm}
\narrowtext
	One technologically important
way to fluidize a granular medium like sand is
to vibrate the walls containing the system.
	Basic issues include the scaling of the kinetic
and potential energy of the grains with parameters such
as the frequency and amplitude of the wall vibration,
and the number of layers of grains.
	These scaling relations are not yet fully
understood for multidimensional systems, despite
progress in simulations\cite{McNamara98},
experiments\cite{Warr95},
and theory\cite{Kumaran98,Huntley98}.

	We consider a granular system in gravity excited
by periodic vertical motion of the container
$z(t) = z_0 cos(\omega t)$.
	When the dimensionless acceleration
$\Gamma = z_0 \omega^2/g$ is larger than one, the granular
system enters a state of internal motion
determined in part by $\Gamma$, the dimensionless
frequency $\overline{\omega} = \omega (d/g)^{1/2}$,
and the aspect ratio of the system.
	Here $g$ is the acceleration of gravity and
$d$ is the grain diameter, so $(d/g)^{1/2}$ is the
time for a grain to fall through its 
radius\cite{McNamara98}.
	For $\overline{\omega} \ll 1$ the container
vibration induces large-scale structures in the granular
medium such as density and surface waves.
	Surface structures are also observed for larger
$\overline{\omega} \approx 1$ in very low aspect-ratio
systems that permit small-wavenumber symmetry
breaking\cite{Pak93,Melo95}.
	Conversely, for $\overline{\omega} \gg 1$ the
container vibration is fast compared with the grain-scale
motion.
	In this limit, the granular medium can assume a
fluidized state analogous to a gas or liquid,
and possibly amenable to theoretical methods developed
for such
systems\cite{Warr95,Kumaran98,Jenkins83,Haff83,Campbell92}.
	The major distinguishing characteristic of a
{\em granular} fluid is the continuous energy loss
to inelastic collisions, which if sufficiently great
can destroy the statistical uniformity of the system
via inelastic collapse\cite{McNamara96}.

	In this paper we report an experimental
study of a three-dimensional granular system
in the fluidized regime
$\overline{\omega} \approx 4$, $\Gamma \le 14$.
	Past numerical and experimental work has
concentrated on one- and two-dimensional
systems\cite{McNamara98,Warr95,Clement91,Luding94,Luding94a,Lee95},
due in part to limited computational resources and the
lack of experimental methods suitable for examining the
interiors of dense, flowing three-dimensional media.
	Recently, nuclear magnetic resonance (NMR) has
emerged as a powerful noninvasive method for studying
three-dimensional granular
media\cite{Nakagawa93,Knight96,Seymour00}.
	In the experiments reported here, NMR imaging
was used to measure the density distribution during
continuous vibration of the container.
	Incoherent grain motion limited image acquisition
time to approximately 1 ms, too short to permit
ordinary two-dimensional imaging.
	Therefore one-dimensional images were acquired
showing the density projected onto the vertical axis
$\rho(z)$.
	We find that $\rho(z)$ deviates strongly
from the exponential form $e^{-Cz}$ expected for
a dilute isothermal gas in gravity\cite{Kumaran98}.
	We focus here on the scaling of the gravitational
potential energy, as measured by the center-of-mass
height, with $\Gamma$ and the number of grains $N$.
	In the present three-dimensional experiments
the scaling is found to deviate even more strongly
from ideal-gas predictions than was found in earlier
two-dimensional studies\cite{Warr95,Luding94a}.

	The granular medium used for these experiments was
composed of mustard seeds, of mean diameter 0.18 cm and
mass 4.0 mg.
	Under magnification, the seeds appeared
roughly spherical with typical eccentricity $\pm 15\%$.
	Seeds are used because their oily centers give strong,
long-lived NMR signals\cite{Nakagawa93,Knight96}.
	The seeds were held in a cylindrical container
formed by a vertical glass tube of inside diameter 0.8 cm with
a flat teflon bottom wall.
	The container was sufficiently tall to prevent
collisions between the seeds and the top wall.
	It was determined visually that $18 \pm 2$ seeds were
required to complete a monolayer in this container.
	The experiments were carried out under ambient
atmospheric pressure.

	A vertical fiberglass tube was used to support the
container at the center of the NMR probe, which was mounted
into a vertical-bore superconducting solenoid.
	The static field was set to $B_0 = 1.00 \text{ T}$
to reduce susceptibility contrast effects in these highly
inhomogeneous samples.
	The lower end of the fiberglass tube was mounted to a
loudspeaker, driven by a function generator and power
amplifier, to provide vertical vibrations of the container.
	A flexible support near the top of the tube permitted
the tube and container to move vertically while avoiding
contact with the stationary NMR probe.

	The waveform and amplitude of vertical motion were
continuously measured by a micromachined accelerometer
(Analog Devices ADXL50) mounted to the fiberglass tube,
close to the NMR probe.
	This accelerometer was initially calibrated against a
more accurate unit, and its output was digitized and fitted.
	Data were taken at two values of the vibration
frequency $\omega/2\pi =$ 50 Hz and 40 Hz.
	Acceleration values were set to within $\pm 0.05g$.

	One-dimensional NMR images were obtained using a
spin-echo sequence.
	Gradient pulses of fixed amplitude 61.5 mT/m were
used to encode vertical position information and an echo
time TE = 1.0 ms was selected.
	Thus, the entire imaging sequence was complete
within 1.5 ms.
	These parameters were chosen to achieve adequate
spatial resolution $\delta z = 800 \text{ }\mu$m within
a time much smaller than the vibration period while holding
signal loss due to incoherent grain motion to an acceptably
small level.
	This loss, measured by the echo size, was greater for
smaller samples and larger vibration amplitudes.
	For a mid-size sample with $N = 50$ the loss was
approximately 10\% at the highest amplitude used.

\begin{figure}
\includegraphics[width=0.9 \linewidth]{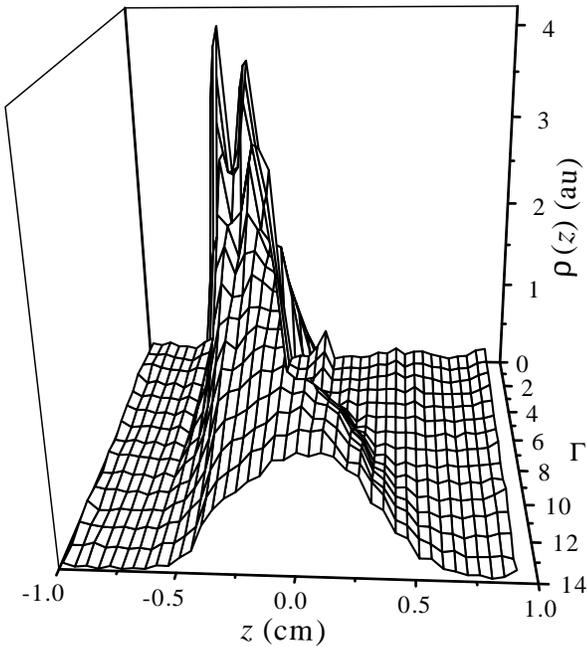}
\vspace{2 mm}
\caption{\label{figrho}
	Density profiles $\rho(z)$ for $N=50$ grains
vibrated at 40 Hz.
	The acceleration of the vibration relative
to gravity $\Gamma$ was varied from 0 to 14.
	The zero of the height axis $z$ is arbitrary,
as are the units of the density axis.
	With no vibration ($\Gamma = 0$), the grains
are organized in distinct layers which appear as
peaks in $\rho(z)$.
	At high values of $\Gamma$, the layer structure
is eradicated and a smooth density profile is observed.}
\end{figure}

	The NMR acquisition was triggered synchronously with
the sample vibration.
	For each value of $N$ and $\Gamma$, data for 100
different trigger delays spread uniformly over the vibration
period were averaged together.
	Thus, the data presented here represent the sample density
{\em averaged over the vibration cycle}.
	To remove effects of rf-field inhomogeneity the data
were normalized to data obtained for a stationary
water-filled tube.

	Figure \ref{figrho} shows the vertical density profile
as a function of $\Gamma$ for one of the $(N, \omega)$ points
measured.
	With no vibration ($\Gamma = 0$) the profile is deeply
corrugated, reflecting the layering of grains in the
stationary chamber.
	As the vibration amplitude is increased, the layer
structure is gradually smoothed out.
	This smearing is due to relative motion of the grains
and not simply the vertical motion of the chamber, as the
latter has an amplitude
$z_0 = (155 \text{ }\mu\text{m})\Gamma$ that is typically
less than the grain diameter.

	Even at the highest vibration amplitude, it is clear
from Fig. \ref{figrho} that the density profile deviates
strongly from the exponential form predicted for an isothermal
ideal gas.
	Although the density tail at the upper (free) surface
might reasonably be fit to an exponential, the density levels
off and then decreases as the lower surface is approached.
	This agrees qualitatively with features observed in
experiments and simulations for two-dimensional
systems\cite{Warr95,Luding94a}.

\begin{figure}
\includegraphics[width=0.9 \linewidth]{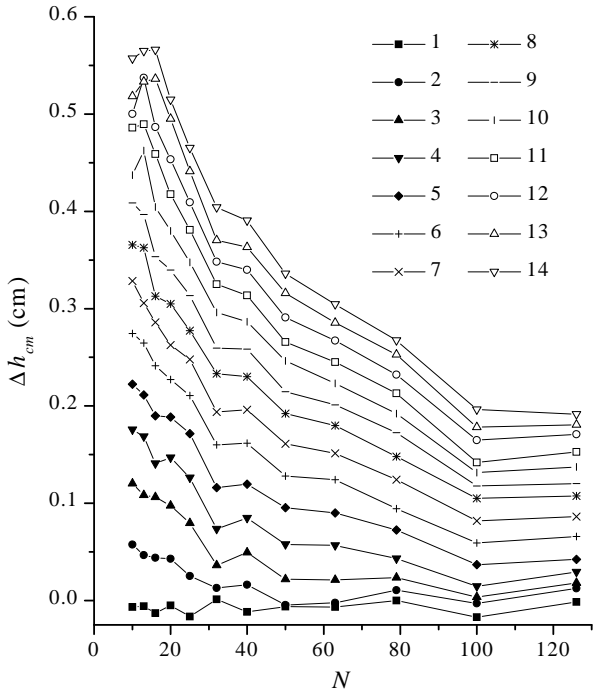}
\vspace{2 mm}
\caption{\label{figraw}
	Rise in the center of mass of the grains relative
to the unvibrated system, as a function of number of
grains $N$ and acceleration $\Gamma$.
	The legend gives the $\Gamma$ value for each
set of symbols.
	This data was taken with 50 Hz vibration frequency.
} \end{figure}

	To quantify the gravitational potential energy of
the system, we have computed the center-of-mass height
$h_{cm} = (\int z \rho(z)dz)/(\int \rho(z)dz)$.
	For each $N$, the limits of integration were kept fixed
as $\Gamma$ was varied to avoid introducing a systematic bias.
	Then the rise in center-of-mass height $\Delta h_{cm}$
was computed by subtracting the $\Gamma = 0$ value of $h_{cm}$
for the same $N$.
	Figure \ref{figraw} shows $\Delta h_{cm}(\Gamma,N)$
for 50 Hz vibration frequency.
	The time-average height of the container is independent
of vibration amplitude, so positive $\Delta h_{cm}$
indicates an upwards displacement of the grains relative to
the average container position.

\begin{figure}
\includegraphics[width=0.9 \linewidth]{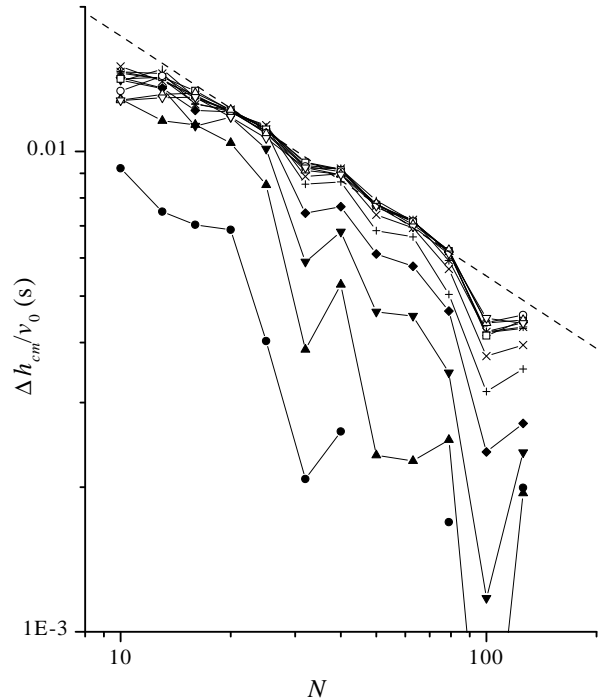}
\vspace{2 mm}
\caption{\label{figscale}
	The data of Fig. 2, divided by the vibration
velocity $v_0$ and plotted on log axes.
	The symbols are the same as in Fig. 2.
	The eight curves for $\Gamma$ = 7-14 lie nearly
on top of one another, implying that $\Delta h_{cm}$
scales as $v_0^\alpha$ with $\alpha \approx 1$.
	The dashed line is the function
$(0.055 \text{ s})/N^{1/2}$.
} \end{figure}

	Small systematic dips can be noted in
Fig. \ref{figraw} for specific values of $N$, for example
$N = 32$ and $N = 100$.
	These do not reflect special states of the fluidized
$(\Gamma \gg 1)$ state for these $N$ values, but rather
variations in the static $(\Gamma = 0)$ packing of the
particles which affect the subtraction used to compute
$\Delta h_{cm}$.
	These dips could be removed from Figs. \ref{figraw}
and \ref{figscale} by assuming a power-law form for
$\Delta h_{cm}$ and fitting to determine the
$\Gamma = 0$ limit of the {\em fluidized} state.
	We prefer to display the data after subtracting the
{\em measured} $\Gamma = 0$ $h_{cm}$ to avoid prejudging
the functional form of $\Delta h_{cm}(\Gamma,N)$.

	To ascertain that functional form, we have plotted
the $\Delta h_{cm}$ data scaled by various powers of the
drive amplitude $z_0$ and frequency $\omega$.
	Figure \ref{figscale} shows the best data collapse
obtained in this way, which is found to be
$\Delta h_{cm}/v_0$.
	Here $v_0 = z_0 \omega$ is the velocity amplitude
of the vibration.
	In this plot, the data for low acceleration
$\Gamma \le 6$ do not collapse while the data for high
acceleration $\Gamma >6$ all collapse onto a single
function of $N$.
	The collapse is significantly better for
$\Delta h_{cm}/v_0$ than it is for $\Delta h_{cm}/v_0^{0.5}$
or $\Delta h_{cm}/v_0^{1.5}$, from which we conclude that
$\Delta h_{cm}$ scales as $v_0^\alpha$ with
$\alpha = 1.0 \pm 0.2$.
	As shown by the dashed line on Fig. \ref{figscale},
the variation of $\Delta h_{cm}$ with $N$ at high
$\Gamma$ is approximately $N^{-1/2}$.
	From plots like this for a range of exponents we
ascertain that $\Delta h_{cm}/v_0 \propto N^{-\beta}$
with $\beta = 0.5 \pm 0.1$.

	A data set at fixed vibration frequency as shown in
Fig. \ref{figscale} can be used to determine the scaling
with $z_0$ but not with $\omega$.
	When the data at vibration frequency 40 Hz are added
to the scaling plot it appears that velocity $z_0 \omega$
is a better scaling variable than is the amplitude $z_0$
or the acceleration $z_0\omega^2$.
	However, due to the limited variation of $\omega$ this
conclusion is not firmly established by our experiments.

	The simplest model for a vibro-fluidized granular
medium is the elastic hard-sphere gas.
	In this model the mean grain velocity is proportional
to the container vibration velocity, and the potential
energy is proportional to the kinetic energy hence
$\Delta h_{cm} \propto v_0^2$.
	This velocity scaling has been observed in experiments
and simulations of {\em one-dimensional} granular systems
(columns of nearly elastic beads)\cite{Luding94}.
	{\em Two-dimensional} experiments\cite{Warr95} and
simulations\cite{Luding94a} found 
$\Delta h_{cm} \propto v_0^\alpha$ with significantly
smaller $\alpha = \text{1.3-1.5}$, a result that is only
partially understood\cite{McNamara98,Huntley98}.
	From a phenomenological viewpoint, our result
$\alpha \approx 1$ in three dimensions appears to continue
the trend with increasing dimensionality away from elastic
kinetic theory predictions.

	For a less superficial understanding we must consider
the reasons for deviations from ideal-gas theory that have
been proposed.
	The steady state of the granular system is determined
by a balance between work done on the grains by collisions
with the vibrating container bottom, and energy lost due to
inelastic collisions between the grains\cite{McNamara98}.
	(Other loss mechanisms discussed in the literature but
not considered here include collisions with the side walls
and viscous damping due to interstitial
gas\cite{Kumaran98}.)

	We first discuss the power $P_i$ that is fed into
the translational degrees of freedom by collisions with
the vibrating bottom wall.
	An important dimensionless parameter is the ratio of
the mean random grain velocity $v_g$ to the characteristic
wall vibration velocity $v_0$.
	For nearly elastic systems (restitution coefficient
close to one) with a small number of particles per unit
wall area, it is possible to have $v_g/v_0 \gtrsim 1$.
	Conversely, less elastic systems (including
our experimental system) generally have $v_g/v_0 \ll 1$
in the uniformly fluidized state at large
$\overline{\omega}$.

 	Collisions between the grains and the vibrating
bottom wall always transfer upward momentum to the
grains, but only do positive work on the grains when the
wall is moving upward.
	When $v_g/v_0 \ll 1$ the grains are unable to catch
up with the wall on its downward stroke.
	In this case, equating the momentum transfer per unit
time with the sample weight $\propto N$ gives the rigorous
result $P_i \propto v_0N$ \cite{McNamara97}.
	The same result obtains for an asymmetric sawtooth
vibration waveform (infinitely fast downward stroke)
for any value of $v_g/v_0$.
	Thus the symmetry of the drive waveform is only
significant when $v_g/v_0 \gtrsim 1$ \cite{McNamara97}.
	For a symmetric waveform, it has been suggested and
verified for two-dimensional simulations that
$P_i \propto v_0Nf(v_g/v_0)$, where $f(0) = 1$ and $f(x)$
is a decreasing function of $x$ \cite{McNamara98}.
	McNamara and Luding\cite{McNamara98} fit their
simulation data for symmetric drive waveforms to
$f(x)=e^{-Cx}$.
	We suggest their data may also be approximately
represented by $f(x)=1/(1+x)$.
	For large $v_g/v_0$, this gives
$P_i \propto v_0^2N/v_g$, a result that has been derived
directly from kinetic theory for symmetric drive
waveforms in the nearly elastic limit\cite{Kumaran98}.

	For a dilute, nearly elastic, isothermal system kinetic
theory predicts that the density decays exponentially with
height and the power lost to collisions
$P_c \propto N^2v_g$ \cite{Kumaran98}.
	Here isothermal denotes uniform granular temperature
$T \propto v_g^2$.
	In this limit potential energy is strictly proportional
to kinetic energy, $\Delta h_{cm} \propto T$.
	Setting $P_c = P_i$, for $v_g/v_0 \ll 1$ or a sawtooth
waveform the granular temperature scales as
$T \propto v_0^2/N^2$.
	For symmetric waveforms with $v_g/v_0 \gg 1$ the
scaling is $T \propto v_0^2/N$ \cite{Kumaran98}.
	The latter corresponds well to experiments and
simulations for one-dimensional systems\cite{Luding94},
but not for two-dimensional systems\cite{Warr95,Luding94a}
or the three-dimensional experiment reported here.

	Huntley\cite{Huntley98} has suggested that correlations
between grain motions at high density reduce the collision
frequency below the kinetic-theory result for an ideal gas.
	The power lost to collisions is found to scale
as $P_c \propto N^{3/2}v_g^2$.
	Combining this with the $v_g/v_0 \gg 1$ result for
$P_i$, Huntley found $T \propto v_0^{4/3}/N^{1/3}$.
	This is much closer to the experimental and simulation
results for two-dimensional systems than is the unmodified
kinetic theory prediction.
	If Huntley's expression for $P_c$ is set equal to $P_i$
for the $v_g/v_0 \ll 1$ condition appropriate to the present
experiments, we obtain $T \propto v_0/N^{1/2}$.
	This agrees well with the scaling seen in our
experiments, suggesting that the degree of inelasticity may
be the controlling factor for scaling exponents rather than
the system dimension.

	Recently, McNamara and Luding\cite{McNamara98}
carried out a series of two-dimensional simulations
in which $P_i$, $P_c$, $T$, and $\Delta h_{cm}$ were measured
as the relevant dimensionless parameters were varied.
	As expected, they found
$\Delta h_{cm} \propto T \propto v_0^2$ in the limit of
very low density that occurs at high drive velocity $v_0$.
	For lower $v_0$, they found a crossover to
$T \propto v_0^{3/2}$ which they traced to a reduction of
$P_c$ below its kinetic-theory value when $\Delta h_{cm}$
is small.
	As the detailed mechanism for this reduction is
unknown, it is not clear how these results can be applied
to the present three-dimensional experiments.
	McNamara and Luding also found that
$\Delta h_{cm} \propto T$ is not well obeyed at high
densities. 

	The trend found in Ref. \cite{McNamara98} of
reduced power loss to collisions at high density agrees
with Huntley's suggestion, although the functional
form may be different.
	As the power input $P_i$ appears reasonbly well
understood, we also interpret our experiments as indicating
reduced power loss to collisions $P_c$ for dense granular
systems in three dimensions, as compared with ideal-gas
kinetic theory predictions.
	
	We thank N. Menon and R. A. Guyer for useful
conversations.  This work was supported by NSF grant
DMR 9501171.


\begin{references}

\bibitem{McNamara98} S. McNamara and S. Luding, Phys. Rev.
E {\bf 58}, 813 (1998).

\bibitem{Warr95} S. Warr, J. Huntley, and G. T. H. Jacques,
Phys. Rev. E {\bf 52}, 5583 (1995).

\bibitem{Kumaran98} V. Kumaran, Phys. Rev. E {\bf 57},
5660 (1998).

\bibitem{Huntley98} J. M. Huntley, Phys. Rev. E {\bf 58},
5168 (1998).

\bibitem{Pak93} H. K. Pak and R. P. Behringer, Phys. Rev.
Lett. {\bf 71}, 1832 (1993).

\bibitem{Melo95} F. Melo, P. H. Umbanhowar, and H. L.
Swinney, Phys. Rev. Lett. {\bf 75}, 3838 (1995).

\bibitem{Jenkins83} J. T. Jenkins and S. B. Savage, J.
Fluid Mech. {\bf 130}, 187 (1983).

\bibitem{Haff83} P. K. Haff, J. Fluid Mech. {\bf 134},
401 (1983).

\bibitem{Campbell92} C. S. Campbell, Ann. Rev. Fluid
Mech. {\bf 22}, 57 (1992).

\bibitem{McNamara96} S. McNamara and W. R. Young,
Phys. Rev. E {\bf 53}, 5089 (1996).

\bibitem{Clement91} E. Cl\'{e}ment and J. Rajchenbach,
Europhys. Lett. {\bf 16}, 133 (1991).

\bibitem{Luding94} S. Luding, E. Cl\'{e}ment, A. Blumen,
J. Rachenbach, and J. Duran, Phys. Rev. E {\bf 49}, 1634
(1994).

\bibitem{Luding94a} S. Luding, H. J. Herrmann, and
A. Blumen, Phys. Rev. E {\bf 50}, 3100 (1994).

\bibitem{Lee95} J. Lee, Physica A {\bf 219}, 305 (1995).

\bibitem{Nakagawa93} M. Nakagawa, S. A. Altobelli,
A. Caprihan, E. Fukushima, and E.-K. Jeong, Exp.
Fluids {\bf 16}, 54 (1993).

\bibitem{Knight96} J. B. Knight, E. E. Ehrichs, V. Yu.
Kuperman, J. K. Flint, H. M. Jaeger, and S. R. Nagel,
Phys. Rev. E {\bf 54}, 5726 (1996).

\bibitem{Seymour00} J. D. Seymour, A. Caprihan, S. A.
Altobelli, and E. Fukushima, Phys. Rev. Lett.
{\bf 84}, 266 (2000).

\bibitem{McNamara97} S. McNamara and J.-L. Barrat,
Phys. Rev. E {\bf 55}, 7767 (1997).

\end{references}
\end{document}